\documentclass[11pt]{article}
\usepackage{epsf}
\newcommand{\eps}{\epsilon}

\newcommand{\sinc}{{\rm sinc}}

\newcommand{\bea}{\begin{eqnarray}}

\newcommand{\eea}{\end{eqnarray}}

\begin{document}


\title{Probability Distributions of Random Electromagnetic Fields in the Presence of a Semi-Infinite Isotropic Medium}

\author{
{\it Luk R. Arnaut}\\
\\
National Physical Laboratory\\
Division of Enabling Metrology\\
Hampton Road,
Teddington,
Middlesex TW11 0LW\\
United Kingdom\\
Fax: +44-20-8943 7176, e-mail: luk.arnaut@npl.co.uk
}


\maketitle

%
%


\begin{abstract}
Using a TE/TM decomposition for an angular plane-wave spectrum of free random electromagnetic waves and matched boundary conditions, we derive the probability density function for the energy density of the vector electric field in the presence of a semi-infinite isotropic medium.
The theoretical analysis is illustrated with calculations and results for good electric conductors and for a lossless dielectric half-space. The influence of the permittivity and conductivity on the intensity, random polarization, statistical distribution and standard deviation of the field is investigated, both for incident plus reflected fields and for refracted fields. External refraction is found to result in compression of the fluctuations of the random field.
\end{abstract}

\section{Introduction}
Complex electromagnetic (EM) environments are characterized by the fact EM fields behave as random or quasi-random quantities. They can be studied in an efficient manner with the aid of statistical electromagnetics methods.
With regard to the interaction of random fields with their environment, a fundamental problem of interest is the evolution of statistical properties of the field upon propagation through stratified media, including reflection and refraction at their interfaces. 
For a perfectly electrically conducting (PEC) surface, previous studies of the average value [Dunn, 1990], standard deviation and first-order probability density function (pdf) [Arnaut and West, 2006] of the electric and magnetic energy densities have demonstrated a direction-dependent damped oscillatory behaviour of their average value and standard deviation as a function of the distance of the point of evaluation to the interface. This behaviour is a consequence of the interference between incident and reflected fields. As a result, unlike for deterministic waves, a boundary zone exists for random fields adjacent to the PEC surface, in which the statistical field properties are inhomogeneous and fundamentally different from those at larger (theoretically infinite) distance. Further insights that were gained from these studies pertain to the statistical anisotropy and polarization state of the field within the boundary zone and, for vector fields, the transitions of the pdf of the energy density from one- or two-dimensionally confined random fields at the interface to fully developed three-dimensional random fields at large electrical distances. 
In addition, spatial correlation functions have been obtained previously for unbounded [Bourret, 1960], [Sarfatt, 1963], [Eckhardt {\it et al.}], [Mehta and Wolf, 1964], [Hill and Ladbury, 2002] and single-interface [Arnaut, 2006a], [Arnaut 2006b] vector EM fields that elucidate the spatial structure of random fields via their two-point coherence properties.

In the present paper, the methods and results for statistical properties of random fields near a PEC surface are extended to a magneto-dielectric isotropic semi-infinite medium. Having analyzed the second-order spatial coherence and correlation properties for an impedance boundary in [Arnaut, 2006b], here we are again concerned with first-order statistical, i.e., local distributional properties only. Based on previous results for nonlocal spatial coherencies of the electric field $\langle E_\alpha({\bf r}_1) E^*_\beta({\bf r}_2) \rangle$ ($\alpha,\beta=x,y,z$), the polarization coefficient and pdf for the local energy density are determined.
Because of the single interface and isotropy of the medium, the polarization coefficient is degenerate, whence the pdfs are one-parameter compound exponential (CE-1) distributions [Arnaut, 2002], [Arnaut and West, 2006]. However, unlike for a PEC medium, the angular spectra of reflected and refracted random fields exhibit directivity because reflection and transmission coefficients of plane waves for a magneto-dielectric semi-infinite medium depend on the wave polarization and angle of incidence. 
We shall confine the analysis and results to the electric field ${\bf E}$; corresponding results for the magnetic field follow without difficulty. Since we express results in terms of Fresnel reflection and transmission coefficients for an isotropic half-space, analogous results for multilayer strata are easily obtained from the listed integral expressions, on substituting with the appropriate coefficients.
In particular, probability distributions of reflected and transmitted fields on either side of a single layer of finite thickness are easily computed.

\section{Field coherencies}

Since random EM fields are spatially and temporally incoherent, their fundamental statistical quantity is the energy density -- rather than the field itself or its magnitude -- which is obtained from the average field intensities $\langle |E_\alpha({\bf r})|^2 \rangle$ of the Cartesian complex field components $E_\alpha$. These intensities represent self-coherencies, which are special cases of mutual coherencies [Mehta and Wolf, 1964]. The latter were derived in [Arnaut, 2006b] for the present configuration. Here, the fact that no separate transverse locations need to be considered in order to obtain first-order statistics considerably simplifies the calculations.

\subsection{Incident plus reflected random field}

\begin{figure}[htb] 
\begin{center} \begin{tabular}{c}
\ \epsfysize=8cm \epsfbox{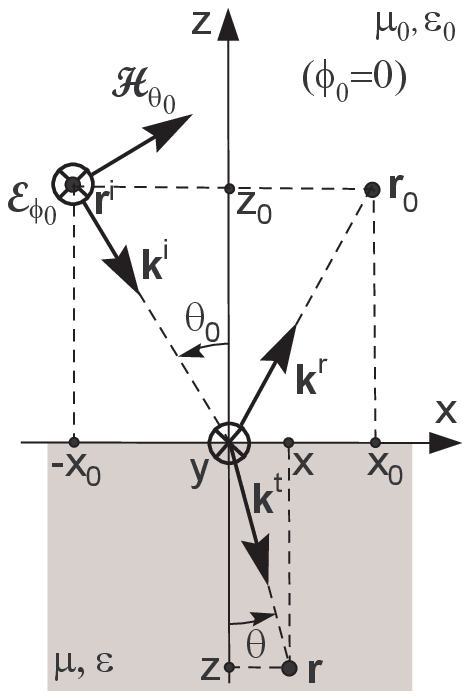}\ \\
\end{tabular}
\end{center}
{\bf \caption{\label{fig:coordTETM}
Coordinate system and local plane of incidence ($\phi_0=0$, ${\bf 1}_\phi={\bf 1}_y$) for single TE wave component reflected and refracted by a semi-infinite isotropic medium.
}}
\end{figure}

We consider a semi-infinite isotropic medium with permittivity $\epsilon$ and permeability $\mu$ occupying the half-space $z \leq 0$ (Fig. \ref{fig:coordTETM}). The incident random field at ${\bf r}$  in the region characterized by $z>0$ is represented by a statistical ensemble (random angular spectrum) of time-harmonic plane waves [Whittaker, 1902], [Booker and Clemmow, 1950], [Hill, 1998]:
\bea
{\bf E}^i({\bf r}) = \frac{1}{2\pi} \int\int_{\Omega_0}{\bf \cal E}^i (\Omega ) \exp (- j {\bf k}^i \cdot {\bf r} ) {\rm d}\Omega,
\label{eq:angularspectrum}
\eea
in which an $\exp(j\omega t)$ time dependence has been assumed and suppressed.
This incident field ${\bf E}^i$ is assumed to be ideal random, i.e., any three complex Cartesian components, in particular ($E^i_x, E^i_y, E^i_z$), are mutually independent and exhibit identical circular centered Gauss normal distributions.
The direction of incidence for each plane-wave component (${\bf \cal E}^i, {\bf \cal H}^i, {\bf k}^i)$ is arbitrary within the upper hemisphere ($\Omega_0 = 2 \pi$ sr) and is specified by azimuthal ($\phi_0$) and elevational ($\theta_0$) angles.
Since the medium is deterministic, the incident and reflected fields for each individual plane wave are mutually {\em coherent}, despite being individually random. Hence, their recombination in the region $z > 0$ is governed by superposition of {\em fields}, rather than energy densities. 
The boundary condition does not affect the correlation between field components on either side of the boundary, because ${\bf \cal E}^i$ is itself random. As a result, the field components remain mutually independent in the vicinity of the boundary.
Since the Cartesian components of the incident field are circular Gaussians and because the medium is linear, the incident plus reflected vector field and its components are also circular Gauss normal with zero mean, but now with different standard deviations from those of the incident field owing to the boundary conditions. Thus, the three Cartesian components no longer exhibit identical parameters for their distributions. 

Following [Dunn, 1990] and [Arnaut, 2006a], we perform a TE/TM decomposition for each plane wave component of the angular spectrum with respect to its associated random plane of incidence $ok^iz$, i.e., $\phi=\phi_0$. 
As is well known, TE and TM polarizations constitute uncoupled eigenpolarizations for stratified media, hence the polarization of the outgoing wave is completely determined by that of the the incoming wave. As a result, the TE and TM contributions to the overall plane-wave spectrum can be calculated independently. 
Ensemble averaging then yields the TE and TM energy contents of the random field.
%
%
Specifically, for the TE components incident at an angle $\theta_0$ and with electric field 
\bea
{\bf \cal E}^i_\perp \exp \left ( - j {\bf k}^i \cdot {\bf r} \right ) 
&=&
{\cal E}_0 \cos \psi 
\exp \left ( j k_0 x \sin \theta_0 \right ) \nonumber\\
&~&\times
\exp \left ( j k_0 z \cos \theta_0 \right ) 
{\bf 1}_y,
\eea
the incident plus reflected electric field at ${\bf r}$ is
\bea
{\cal E}_\perp \exp (-j \bf{k} \cdot \bf{r}) 
&=& {\cal E}_{0} \exp \left ( - j k_0 \varrho \sin\theta_0 \right ) 
\nonumber\\ &~& \times
\left \{ \left [ 1 + \Gamma_\perp (\theta_0) \right ] \cos \left ( k_0 z \cos \theta_0 \right ) \right.\nonumber\\
&~& \left.
     + j \left [ 1 - \Gamma_\perp (\theta_0) \right ] \sin \left ( k_0 z \cos \theta_0 \right ) 
\right \},
\label{eq:EtyTE}
\eea
with $\varrho \stackrel{\Delta}{=} x \cos \phi_0 + y \sin \phi_0$, and where $\Gamma_\perp(\theta_0)$ represents the Fresnel reflection coefficient for TE waves for a semi-infinite isotropic medium.
Upon substitution of (\ref{eq:EtyTE}) into an expression similar to (\ref{eq:angularspectrum}) for ${\bf E}_\perp$, followed by unfolding and integration, the associated average field intensity is
\bea
\langle |E_y ({\bf r})|^2 \rangle = I_{y1} + I_{y2} + I_{y3} + I_{y4},
\label{eq:TE}
\eea
in which $\langle \cdot \rangle$ denotes ensemble averaging of the plane-wave spectrum, with
\bea
I_{y1} &=& 2C
\int^1_0
\left | 1 - \Gamma_\perp (u ) \right |^2 
\sin^2 \left ( k_0 z u \right )
{\rm d}u,
\label{eq:Iy1TE}\\
I_{y2} &=& 2C
\int^1_0
\left | 1 + \Gamma_\perp (u ) \right |^2 
\cos^2 \left ( k_0 z u \right ) 
{\rm d}u,\\
I_{y3} &=& I^*_{y4} = j 2C
\int^1_0
\left [ 1 - \Gamma_\perp (u ) \right ] \left [ 1 + \Gamma^*_\perp (u ) \right ] \nonumber\\
&~&\times
\sin \left ( k_0 z u \right ) \cos \left ( k_0 z u \right )
{\rm d}u,
\label{eq:Iy34TE}
\eea
where $u\stackrel{\Delta}{=} \cos \theta_0$, $C \stackrel{\Delta}{=} \langle |{\cal E}_0|^2 \rangle/4$, 
and 
\bea
\Gamma_\perp(u) = 
\frac{\eta k u - \eta_0 \sqrt{k^2-k^2_0+k^2_0 u^2}}
     {\eta k u + \eta_0 \sqrt{k^2-k^2_0+k^2_0 u^2}}.
\eea
Similarly, for the TM components with incident electric field
\bea
{\bf \cal E}^i_\parallel \exp \left ( - j {\bf k}^i \cdot {\bf r} \right ) 
&=&
- {\cal E}_0 \sin \psi 
\exp \left ( j k_0 x \sin \theta_0 \right ) \nonumber\\
&~&\times
\exp \left ( j k_0 z \cos \theta_0 \right )\nonumber\\
&~&\times 
\left [ \cos \theta_0 {\bf 1}_x + \sin \theta_0 {\bf 1}_z \right ],
\eea
we obtain
\bea
\langle |E_x ({\bf r})|^2 \rangle = \sum^4_{\ell=1} I_{x \ell}, \hspace{0.7cm} \langle |E_z ({\bf r})|^2 \rangle = \sum^4_{\ell=1} I_{z \ell},
\eea
with
\bea
I_{x1} &=& 2C
\int^1_0
\left | 1 - \Gamma_\parallel (u ) \right |^2 u^2
\sin^2 \left ( k_0 z u \right ) 
{\rm d}u,
\label{eq:Ix1TM}\\
I_{x2} &=& 2C
\int^1_0
\left | 1 + \Gamma_\parallel (u ) \right |^2 u^2
\cos^2 \left ( k_0 z u \right ) 
{\rm d}u,\\
I_{x3} &=& I^*_{x4} = j 2C
\int^1_0
\left [ 1 - \Gamma_\parallel (u ) \right ] \left [ 1 + \Gamma^*_\parallel (u ) \right ] \nonumber\\
&~& \times u^2 
\sin \left ( k_0 z u \right ) \cos \left ( k_0 z u \right )
{\rm d}u,
\label{eq:Ix4TM}
\eea
\bea
I_{z1} &=& 2C
\int^1_0
\left | 1 + \Gamma_\parallel (u ) \right |^2 (1-u^2)
\sin^2 \left ( k_0 z u \right ) 
{\rm d}u, 
\\
I_{z2} &=& 2C
\int^1_0
\left | 1 - \Gamma_\parallel (u ) \right |^2 (1-u^2)
\cos^2 \left ( k_0 z u \right ) 
{\rm d}u, \label{eq:Iz1TM} 
\\
I_{z3} &=& I^*_{z4} = j 2C
\int^1_0
\left [ 1 + \Gamma_\parallel (u ) \right ] \left [ 1 - \Gamma^*_\parallel (u ) \right ] 
\nonumber\\
&~& \times (1-u^2)
\sin \left ( k_0 z u \right ) \cos \left ( k_0 z u \right )
{\rm d}u,
\label{eq:Iz3TM}
\eea
where
\bea
\Gamma_\parallel(u) =
\frac{\eta \sqrt{k^2-k^2_0+k^2_0 u^2} - \eta_0 k u}
     {\eta \sqrt{k^2-k^2_0+k^2_0 u^2} + \eta_0 k u}.
\eea
For the tangential and overall (i.e., incident plus reflected) vector field,
\bea
\langle |E_t ({\bf r})|^2 \rangle =
\sum_{\alpha=x,y} \sum^4_{\ell=1} I_{\alpha\ell}
\eea
and
\bea
\langle |E ({\bf r})|^2 \rangle =
\sum_{\alpha=x,y,z} \sum^4_{\ell=1} I_{\alpha\ell},
\eea
respectively. For a medium for which $|\mu_r / \epsilon_r | \rightarrow 0$, only the terms $I_{\alpha1}$ remain nonzero. Conversely, for $|\epsilon_r / \mu_r | \rightarrow 0$, only the terms $I_{\alpha2}$ survive.

\subsection{Refracted random field}
For the field refracted across the boundary,
\bea
\langle |E_y|^2 \rangle &=& 
2C \int^1_0 |T_\perp(u)|^2 
{\rm d}u,
\label{eq:Eysqintegral}\\
\langle |E_x|^2 \rangle &=& 
2C \int^1_0 |T_\parallel(u)|^2 
\left [ 
1 - 
\left ( \frac{k_0}{k} \right )^2 + 
\left ( \frac{k_0}{k} \right )^2 u^2 
\right ]
{\rm d}u,
\label{eq:Exsqintegral}\\
\langle |E_z|^2\rangle &=& 
2C \int^1_0 |T_\parallel(u)|^2 
\left ( \frac{k_0}{k} \right )^2 
\left ( 1 - u^2 \right )
{\rm d}u,
\label{eq:Ezsqintegral}
\eea
with the TE and TM transmission coefficients given by
\bea
T_\perp(u) = \frac{2 \eta k u}{\eta k u + \eta_0 \sqrt{k^2 - k^2_0 + k^2_0 u^2} }
\eea
and
\bea
T_{\parallel}(u) = \frac{2 \eta k u}{\eta_0 k u + \eta \sqrt{k^2 - k^2_0 + k^2_0 u^2} },
\eea
respectively. Thus, unlike for the incident plus reflected field, the intensity of the refracted field is homogeneous due to the absence of interference in the region $z>0$.

\section{Energy density distribution}
For a single-interface configuration, the pdf of the electric energy density $\langle S_e \rangle = \epsilon_{(0)} \langle |E|^2 \rangle /2$ can be calculated based on knowledge of the polarization coefficient $P_{e,i3}$ for the statistically uniaxial electric field [Arnaut and West, 2006]. For the vector tangential field, the incoherent superposition $\langle |E_t|^2 \rangle = \langle |E_x|^2 \rangle + \langle |E_y|^2 \rangle$ holds, because ${\cal E}_x$ and ${\cal E}_y$ belong to mutually orthogonal modes. Hence,
\bea
P_{e,13} = P_{e,23} 
&=&
\frac{1}{2} - \frac{3 \langle |E_z(k_0z)|^2 \rangle}{2 \langle |E(k_0z)|^2 \rangle},
\eea
in which $E_z\equiv E_3$, and
\bea
\langle |E(k_0z)|^2 \rangle
&=&
\sum_{\alpha=x,y,z}
\langle |E_\alpha(k_0z)|^2 \rangle \\
&=& 2C \int^1_0 \left [ |T_\perp(u)|^2 + |T_\parallel(u)|^2 \right ]
{\rm d}u.
\eea
The associated CE-1 pdf of $S_e$ follows as [Arnaut, 2002], [Arnaut and West, 2006]
\bea
f_{S_e} (s_e) &=& 
\gamma_1 s_e \exp \left ( - \alpha_1 s_e \right ) + 
\gamma_2     \exp \left ( - \alpha_2 s_e \right ) 
+ 
\gamma_3     \exp \left ( - \alpha_3 s_e \right ),
\label{eq:CEnearwall}
\eea
where 
\bea
\alpha_1 =\alpha_2 
&=&
\frac{3}{ (1+P_{e,13}) \langle S_e \rangle }
=
\frac{2}{ \langle S_e \rangle - \langle S_{e_z} \rangle },
\eea
\bea
\label{eq:alpha1}
\alpha_3 
&=& 
\frac{3}{ (1-2P_{e,13}) \langle S_e \rangle }
= \frac{1}{\langle S_{e_z} \rangle},
\eea
\bea
\gamma_1 &=& \frac{3}{ P_{e,13}(1+P_{e,13}) \langle S_e \rangle^2 }\nonumber\\
&=&
\frac{4}{\left ( \langle S_e \rangle - 3 \langle S_{e_z} \rangle \right ) \left ( \langle S_e \rangle - \langle S_{e_z} \rangle \right )},
\\
\gamma_2 &=& \frac{2 P_{e,13}^2 + P_{e,13} -1 }{3 P_{e,13}^2 (1+P_{e,13}) \langle S_e \rangle  }\nonumber\\
&=& -
\frac{4 \langle S_{e_z} \rangle }
{ \left ( \langle S_e \rangle - 3 \langle S_{e_z} \rangle \right )^2 },
\\
\gamma_3 &=& \frac{4 P_{e,13}^2 - 4 P_{e,13} +1 }{3 P_{e,13}^2 (1-2P_{e,13}) \langle S_e \rangle  }\nonumber\\ 
&=&
\frac{4 \langle S_{e_z} \rangle}{ \left ( \langle S_e \rangle - 3 \langle S_{e_z} \rangle \right )^2 }  = - \gamma_2
\label{eq:gamma3}.
\eea

\section{Special cases}
\subsection{Good electric conductors}
For nonmagnetic good electric conductors, i.e., $\sigma \gg \omega \epsilon_0$, $\epsilon=\epsilon_0$, $\mu=\mu_0$, it follows that $\eta/\eta_0 \simeq \sqrt{\omega \epsilon_0 / (2\sigma)} (1\pm j)$, $\Gamma_\perp(u) \simeq (\eta u -\eta_0)/(\eta u + \eta_0)$ and $\Gamma_\parallel(u) \simeq (\eta -\eta_0 u)/(\eta + \eta_0 u)$ because $\theta \simeq 0$ irrespective of $\theta_0$. As a result, 
(\ref{eq:Iy1TE})--(\ref{eq:Iy34TE}) become
\bea
I_{y1} &=& 4 C \int^1_0 \frac{1-\cos(2k_0zu)}{\left | 1 + \frac{\eta}{\eta_0}u \right |^2}{\rm d}u,\\
I_{y2} &=& 4 C \left | \frac{\eta}{\eta_0} \right |^2 \int^1_0 \frac{1+\cos(2k_0zu)}{\left | 1 + \frac{\eta}{\eta_0}u \right |^2} u^2 {\rm d}u,\\
I_{y3} &=&  I^*_{y4} = j 4 C 
\frac{\eta^*}{\eta_0} 
\int^1_0 
\frac{\sin(2k_0zu)}
     {\left | 1 + \frac{\eta}{\eta_0}u \right |^2}
u
{\rm d}u.
\eea
Throughout the range of integration, $|1+(\eta/\eta_0)u|^2 = [1 + u \sqrt{\omega\epsilon_0/(2\sigma)}]^2 + u^2[\omega\epsilon_0/(2\sigma)] \simeq 1$. Therefore, to good approximation, we have
\bea
\langle |E_y ({\bf r})|^2 \rangle 
&\simeq& 
4 C 
\left \{
\left | \frac{\eta}{\eta_0} \right |^2 \frac{2\cos (2k_0z)}{(2k_0z)^2} + 
\left | \frac{\eta}{\eta_0} \right |^2 \sinc (2k_0z) 
\right. \nonumber\\ &~&~~~\left.
-
\left | \frac{\eta}{\eta_0} \right |^2 \frac{\sinc(2k_0z)}{(2k_0z)^2} 
- 
\sinc \left ( 2k_0z \right ) 
\right. \nonumber\\ &~& \left.
- j \frac{\eta-\eta^*}{\eta_0}
\left [ \frac{\sinc(2k_0z)}{(2k_0z)^2} - \frac{\cos(2k_0z)}{(2k_0z)} \right ]
\right \}.
\label{eq:Iysimple}
\eea
For the TM components, (\ref{eq:Ix1TM})--(\ref{eq:Iz3TM}) specialize to
\bea
I_{x1} &=& 4 C \int^1_0 \frac{1-\cos(2k_0zu)}{\left | u + \frac{\eta}{\eta_0} \right |^2} u^4 {\rm d}u,\\
I_{x2} &=& 4 C \left | \frac{\eta}{\eta_0} \right |^2 \int^1_0 \frac{1+\cos(2k_0zu)}{\left | u + \frac{\eta}{\eta_0} \right |^2} u^2 {\rm d}u,\\
I_{x3} &=& I^*_{x4} = j 4 C \frac{\eta^*}{\eta_0} \int^1_0 
\frac{\sin(2k_0zu)}
     {\left | u + \frac{\eta}{\eta_0} \right |^2} u^3 
{\rm d}u,
\eea
\bea
I_{z1} &=& 4 C \int^1_0 \frac{ 1+\cos(2k_0zu) }{\left | u + \frac{\eta}{\eta_0} \right |^2} \left ( 1-u^2 \right ) u^2 {\rm d}u,\\
I_{z2} &=& 4 C \left | \frac{\eta}{\eta_0} \right |^2 \int^1_0 \frac{ 1-\cos(2k_0zu) }{\left | u + \frac{\eta}{\eta_0} \right |^2} \left ( 1-u^2 \right ) {\rm d}u,
\\
I_{z3} &=& I^*_{z4} = j 4 C 
\frac{\eta}{\eta_0} 
\int^1_0 
\frac{ \sin(2k_0zu)}      
     {\left | u + \frac{\eta}{\eta_0} \right |^2} \left ( 1-u^2 \right ) u
{\rm d}u.
\eea

\begin{figure}[htb] 
\begin{center} \begin{tabular}{c}
\ \epsfxsize=12cm \epsfbox{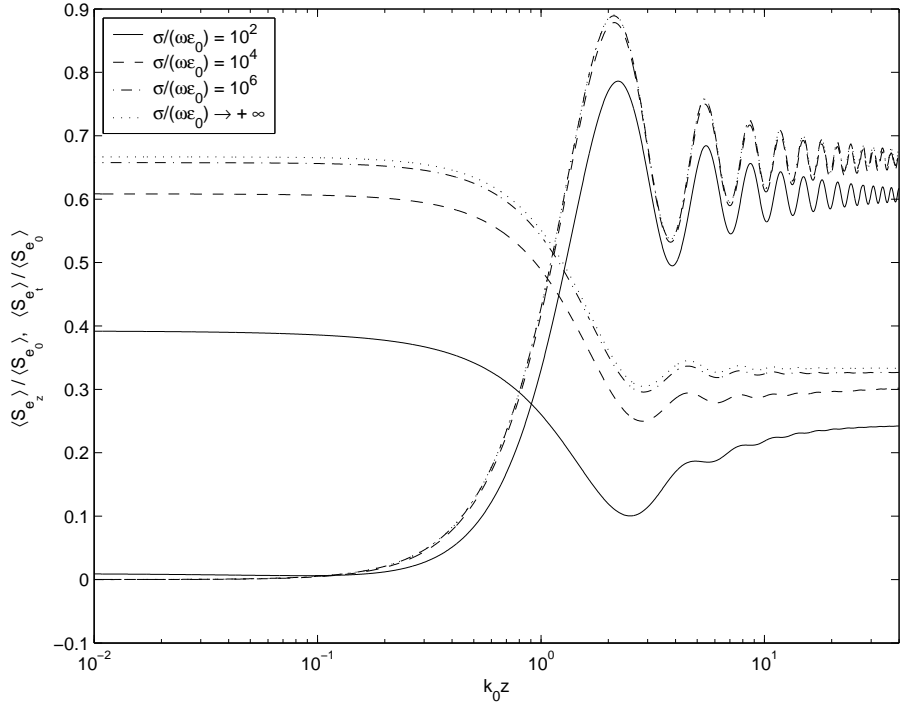}\ \\
\end{tabular}
\end{center}
{\bf \caption{\label{fig:CEnearconduc_avgSeCart_param_sigma}
Average energy densities for normal and tangential components of the incident plus reflected electric field at selected values of $\sigma/(\omega\eps_0)$, normalized with respect to the energy density of the incident vector field, $ \langle S_{e_0} \rangle $. Functions increasing at $k_0 z=1$ represent $\langle S_{e_t} \rangle / \langle S_{e_0} \rangle $; functions decreasing at $k_0 z=1$ represent $\langle S_{e_z}\rangle / \langle S_{e_0} \rangle $.
}}
\end{figure}

Before presenting results for the pdfs, we show in Fig. \ref{fig:CEnearconduc_avgSeCart_param_sigma} the dependencies of
$\langle S_{e_t}\rangle$ and 
$\langle S_{e_z}\rangle$ on $k_0z$ at selected values of $\sigma/(\omega\epsilon_0)$, after normalization by the average electric energy density of the incident total (vector) field, $\langle S_{e_0}\rangle$. 
The sensitivity to variations of $\sigma/(\omega\epsilon_0)$ is seen to be significantly higher for the normal component than for the tangential component. For finite $\sigma/(\omega\epsilon_0)$, the value of $\langle S_{e_t}(k_0z{\rightarrow}0) \rangle/\langle S_{e_0}\rangle$ is close to, but different from zero on account of the EM boundary condition. Both asymptotic values $\langle S_{e_t}(k_0z{\rightarrow}+\infty)\rangle/\langle S_{e_0} \rangle$ and $\langle S_{e_z}(k_0z{\rightarrow}+\infty)\rangle/\langle S_{e_0}\rangle$ are smaller for a finitely conducting boundary than for a PEC boundary, on account of energy dissipation in the former. For intermediate values of $k_0z$, finite conductivity also gives rise to a relatively small positive or negative phase shift in the oscillatory behaviour of $\langle S_{e_z}(k_0z)\rangle/\langle S_{e_0} \rangle$ or $\langle S_{e_t}(k_0z)\rangle/\langle S_{e_0}\rangle$.

\begin{figure}[htb] 
\begin{center} \begin{tabular}{c}
\ \epsfxsize=12cm \epsfbox{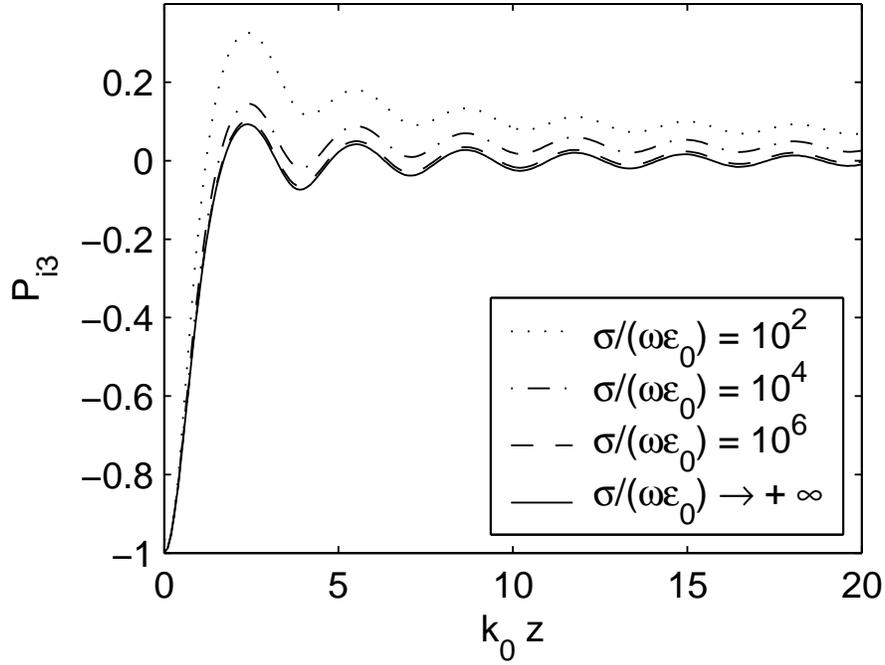}\ \\
\end{tabular}
\end{center}
{\bf \caption{\label{fig:CEnearconduc_Pi3_param_sigma}
Polarization coefficients $P_{i3}$ ($i=1,2$) for the electric energy density of the incident plus reflected field near a conducting medium, at selected values of $\sigma/(\omega\eps_0)$.
}}
\end{figure}

The effect of finite conductivity on the statistical polarization state of the field is shown in Fig. \ref{fig:CEnearconduc_Pi3_param_sigma}. Unlike for a PEC surface, $ P_{i3}(k_0z) \equiv P_{e,i3}(k_0z) $ ($i=1,2$) is no longer oscillating symmetrically with respect to zero, i.e., the random polarization exhibits a conductivity-dependent bias in the normal direction. This bias decreases, on average, with increasing $k_0z$, but persists nevertheless up to asymptotically large distances $k_0z\rightarrow +\infty$.

Figure \ref{fig:CEdistrnearconduc_param_sigma_kz0_0p7854} shows scaled pdfs $f_{S_e}(s_e)$ for selected values of $\sigma/(\omega\eps_0)$ at $k_0z=\pi/4$. At this intermediate distance, the pdfs rapidly approach the asymptotic distribution for $S_e$ near a PEC wall [Arnaut and West, 2006] when $\sigma/(\omega\eps_0)$ is increased. For larger $k_0z$ (not shown), the influence of $\sigma/(\omega\eps_0)$ on the pdf was found to be even weaker.
On the other hand, very close to the boundary ($k_0z \ll 1$), the pdf rapidly approaches a distribution close to the $\chi^2_2$ limit distribution when $\sigma/(\omega\epsilon_0)$ is increased, as demonstrated in Fig. \ref{fig:CEdistrnearconduc_param_sigma_kz0_0p01} for $k_0z=0.01$. 

\begin{figure}[htb] 
\begin{center} \begin{tabular}{c}
\ \epsfxsize=12cm \epsfbox{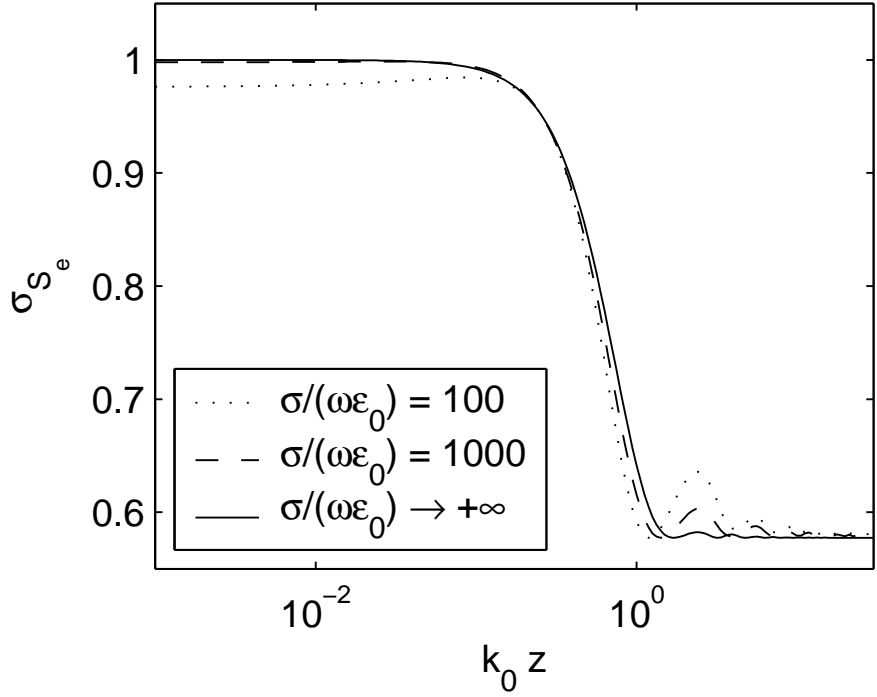}\ \\
(a)\\
\ \epsfxsize=12cm \epsfbox{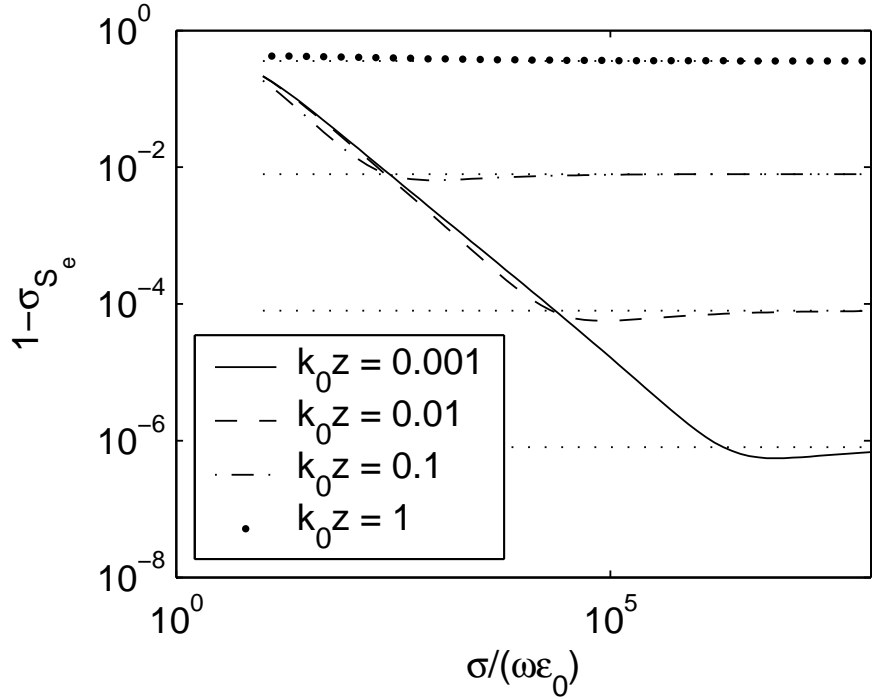}\ \\
(b)\\
\end{tabular}
\end{center}
{\bf \caption{\label{fig:CEnearconduc_std_ifv_kz0_param_sigma}
Std of $S_e$ of the incident plus reflected total field: 
(a) as a function of $k_0z$ for selected values of $\sigma/(\omega\eps_0)$; 
(b) as a function of $\sigma/(\omega\eps_0)$ at selected values of $k_0 z$. Dotted lines in figure (b) represent limit values for a PEC surface at the indicated values of $k_0 z$. 
}}
\end{figure}

As indicated by Fig. \ref{fig:CEnearconduc_std_ifv_kz0_param_sigma}(a), the standard deviation (std) $\sigma_{S_e}$ for the incident plus reflected total field exhibits increasing oscillations for intermediate distances and lower values upon approaching the surface when conductivity is decreased.
The std approaches a location-dependent asymptotic value in the manner shown in Fig. \ref{fig:CEnearconduc_std_ifv_kz0_param_sigma}(b).

\subsection{Lossless dielectric medium}
For a lossless medium, $\eta$, $k$, $T_{\perp,\parallel}$ and $\Gamma_{\perp,\parallel}$ are real-valued. Therefore, $I_{\alpha 3}$ and $I_{\alpha 4}$ are purely imaginary so that only $I_{\alpha 1}$ and $I_{\alpha 2}$ contribute to the intensities $\langle |E_\alpha|^2 \rangle$.
The following explicit expressions are obtained for the energy density of the fields refracted by a lossless dielectric medium with relative permittivity $\epsilon_r\equiv \epsilon/\epsilon_0$:
for TE wave components,
\bea
\langle |E_y({\bf r})|^2 \rangle 
&=& 2 C
\left [
\frac{4}{3\left ( \eps_r - 1 \right )} 
+ \frac{16 \eps^{3/2}_r}{15 \left ( \eps_r -1 \right )} 
\right . \nonumber\\ &~& \left .
- \frac{16 \sqrt{\eps_r-1}}{15} 
- \frac{8 \left ( \eps^{3/2}_r -1 \right )}{5 \left ( \eps_r - 1 \right )^2}
\right ],
\eea
whereas for the TM components, the integrals
\bea
\int^1_0 \left [ T_{\parallel}(u) \right ]^2 {\rm d}u = 2C \sum^5_{i=1} U_i ,
\eea
with
\bea
U_1 &=& \frac{ 4 \eps_r (\eps^2_r-\eps_r+1) - 4 {\eps_r}^{3/2} \left (2 \epsilon_r -1 \right )} { (\eps_r-1)^2 (\eps_r+1)^2 }, \\
U_2 &=& \frac{12 \eps^2_r}{ (\eps_r-1)^{3/2} (\eps_r+1)^2 }, \\
U_3 &=& \frac{ 2\eps_r \left ( 2\eps^2_r + 1 \right )}{ (\eps_r-1)^2 (\eps_r+1)^{5/2}} {\rm ln} \left ( \frac{\sqrt{\eps_r+1}-1}{\sqrt{\eps_r+1}+1} \right ), \\
U_4 &=& \frac{ 2\eps_r \left ( 2\eps^2_r + 1 \right )}{ (\eps_r-1)^2 (\eps_r+1)^{5/2}} 
{\rm ln} \left ( \frac{ \sqrt{ \eps_r+1}+\sqrt{\eps_r} }{ \sqrt{\eps_r+1}-\sqrt{\eps_r} } \right ), \\
U_5 &=& \frac{ 2\eps_r \left ( 2\eps^2_r + 1 \right )}{ (\eps_r-1)^2 (\eps_r+1)^{5/2}} 
{\rm ln} \left ( \frac{ {\eps_r} - \sqrt{ \eps^2_r-1} }{ {\eps_r} + \sqrt{\eps^2_r-1} } \right ), 
\eea
and
\bea
\int^1_0 u^2 \left [ T_{\parallel}(u) \right ]^2 {\rm d}u = 2C \sum^6_{i=1} V_i , 
\eea
with
\bea
V_1 &=& \frac{ 4 \eps_r \left ( \eps^2_r+1 \right ) - 8 \eps^2_r \left [ \eps^{3/2}_r - \left (  \eps_r - 1 \right )^{3/2} \right ] }{3(\eps^2_r-1)^2},\\
V_2 &=& \frac{ 4 \eps_r \left ( 3 \eps^2_r - \eps_r + 1 \right ) - 4 \eps^{3/2}_r \left ( 4 \eps_r - 1 \right ) }{ (\eps_r-1)^2 (\eps_r+1)^3 },\\
V_3 &=& \frac{20 \eps^2_r}{\left ( \eps_r -1 \right )^{3/2} \left ( \eps_r + 1 \right )^3 },\\
V_4 &=& \frac{2 \eps_r ( 4 \eps^2_r + 1 )}{ (\eps_r-1)^2 (\eps_r+1)^{7/2}} {\rm ln} \left ( \frac{\sqrt{\eps_r+1}-1}{\sqrt{\eps_r+1}+1} \right ), \\
V_5 &=& \frac{2 \eps_r \left ( 4 \eps^2_r - \eps_r + 1 \right ) }{(\eps_r-1)^2 (\eps_r+1)^{7/2}} 
{\rm ln} \left ( \frac{ \sqrt{ \eps_r+1}+\sqrt{\eps_r} }{ \sqrt{\eps_r+1}-\sqrt{\eps_r} } \right ),\\
V_6 &=& \frac{2 \eps_r \left ( 4 \eps^2_r - \eps_r + 1 \right ) }{(\eps_r-1)^2 (\eps_r+1)^{7/2}} 
{\rm ln} \left ( \frac{ {\eps_r} - \sqrt{ \eps^2_r-1} }{ {\eps_r} + \sqrt{\eps^2_r-1} } \right )
\eea
can be substituted into (\ref{eq:Exsqintegral}) and (\ref{eq:Ezsqintegral}) to yield $\langle |E_x|^2 \rangle$ and $\langle |E_z|^2 \rangle$ for the refracted field.

\begin{figure}[htb] 
\begin{center} \begin{tabular}{c}
\ \epsfxsize=12cm \epsfbox{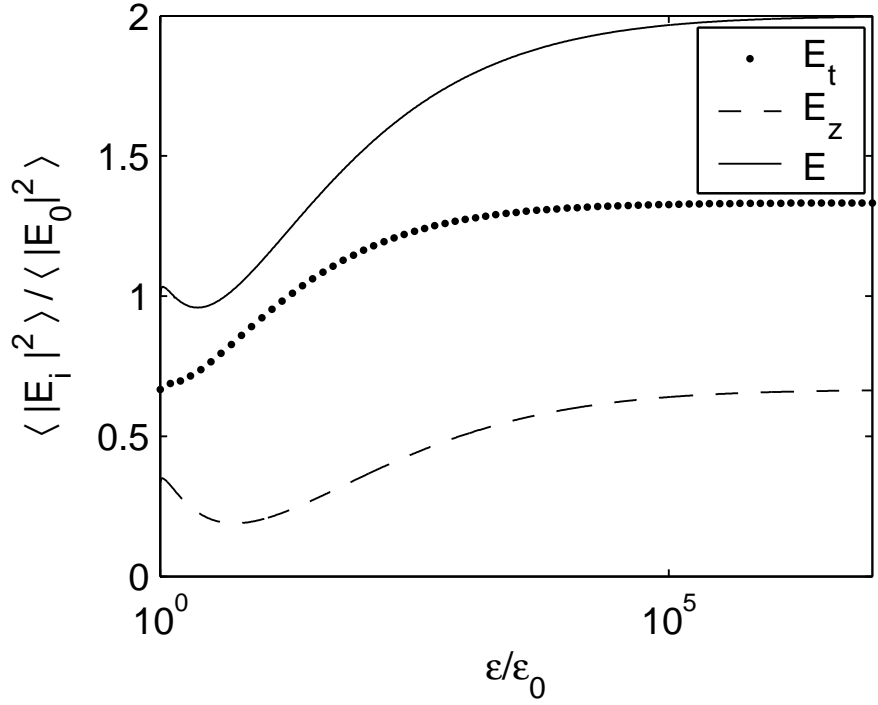}\ \\
(a)\\
\ \epsfxsize=12cm \epsfbox{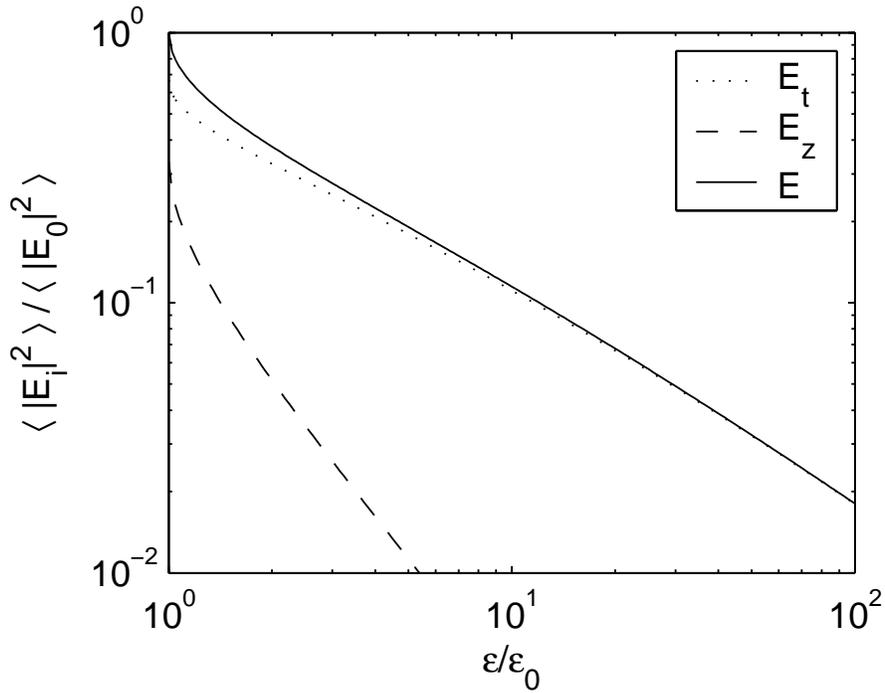}\ \\
(b)\\
\end{tabular}
\end{center}
{\bf \caption{\label{fig:CEneardielec_Salpha_ifv_epsr}
Intensities of tangential and normal electric field components as a function of relative permittivity: (a) for the incident plus reflected field at $k_0z=20\pi$; (b) for the refracted field.
}}
\end{figure}

Figure \ref{fig:CEneardielec_Salpha_ifv_epsr} shows the intensities of the tangential and normal components as a function of permittivity, for the incident plus reflected field at $k_0z=20\pi$, and for the refracted field at any $k_0z$. For the former case [Fig. \ref{fig:CEneardielec_Salpha_ifv_epsr}(a)], the minimum in the curves is a result of reduced TM contributions near the Brewster angle for external refraction ($\eps / \eps_0 > 1$). This effect occurs in both the tangential and normal field components, via $\langle |E_x|^2\rangle$ and $\langle |E_z|^2\rangle$. However, for $\langle |E_t|^2 \rangle$, it causes only a minor dip in the characteristic because of the dominance of the contribution by $\langle |E_y|^2 \rangle$ over that by $\langle |E_x|^2 \rangle$. 
Note the large sensitivity to the permittivity in the vicinity of $\epsilon/\epsilon_0=1$. This sensitivity decreases with increasing $k_0z$. For the normal field, the minimum intensity occurs around $ \epsilon/\epsilon_0\simeq 5.4$. The incident energy density for this component is being exceeded only above relatively high permittivities ($\epsilon/\epsilon_0\sim 70$).
For the refracted field [Fig. \ref{fig:CEneardielec_Salpha_ifv_epsr}(b)], the intensities rapidly decrease with increasing $\epsilon/\epsilon_0$ and exhibit an increasingly dominant tangential contribution.

\begin{figure}[htb] \begin{center} \begin{tabular}{c}
\ \epsfxsize=12cm \epsfbox{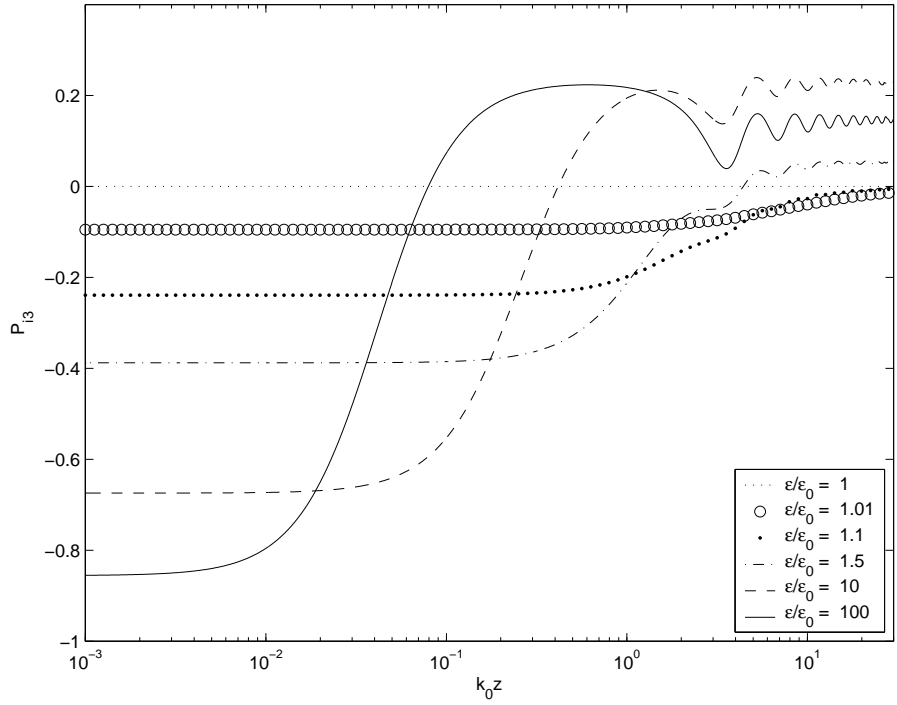}\ \\
(a)
\\
\ \epsfxsize=12cm \epsfbox{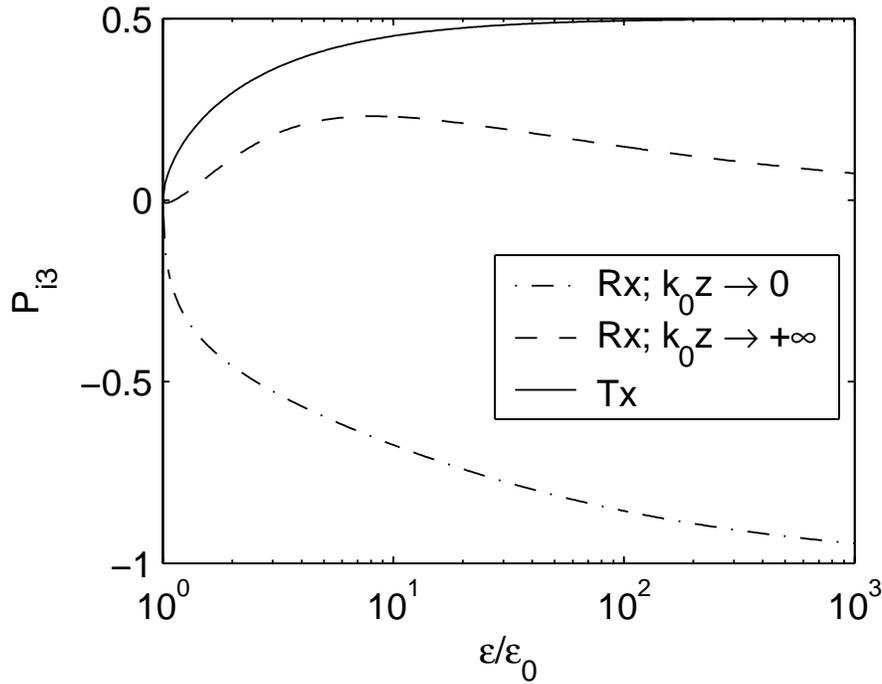}\ \\
(b)
\\
\end{tabular}
\end{center}
{\bf \caption{\label{fig:CEneardielec_Pi3_ifv_epsr}
Polarization coefficients $P_{i3}$ ($i=1,2$) for a lossless dielectric medium: (a) for the incident plus reflected field, as a function of electrical distance from the boundary, at selected values of the permittivity; 
(b) for the incident plus reflected field (Rx; at arbitrarily large or small distance) and for the refracted field (Tx; at any distance) as a function of the relative permittivity.
}}
\end{figure}

Figure \ref{fig:CEneardielec_Pi3_ifv_epsr} shows the polarization coefficient as a function of $\epsilon/\epsilon_0$.
For the incident plus reflected field, an $\eps_r$-dependent threshold distance exists where $P_{i3}$ starts to increase, in an oscillatory manner, toward an ($\eps_r$-dependent) asymptotic value when $k_0z \rightarrow +\infty$. 

\begin{figure}[htb] 
\begin{center} \begin{tabular}{c}
\ \epsfxsize=12cm \epsfbox{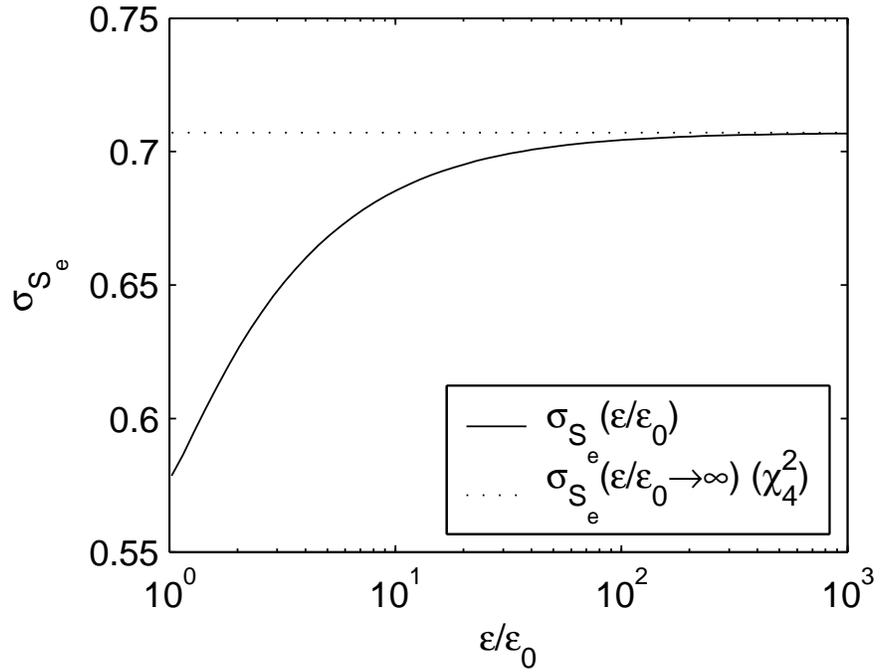}\ \\
\end{tabular}
\end{center}
{\bf \caption{\label{fig:CEneardielec_std_ifv_epsr}
Std of $S_e$ of the refracted field as a function of $\eps_r$ at arbitrary $k_0z$.
}}
\end{figure}

Figure \ref{fig:CEdistrneardielec_Tx_param_epsr} shows $f_{S_e}(s_e)$ for the refracted field at selected values of $\eps / \eps_0$. This pdf is independent of $k_0z$ but evolves from a $\chi^2_6$ distribution for $\eps / \eps_0=1$ to a $\chi^2_4$ distribution for $\eps \rightarrow +\infty$. 
%
From Fig. \ref{fig:CEneardielec_std_ifv_epsr}, $\sigma_{S_e}(\epsilon / \eps_0)$ is seen to increase monotonically from its $\chi^2_6$ value, $1/\sqrt{3}$, at $\epsilon/\epsilon=1$ to its asymptotic $\chi^2_4$ value, $1/\sqrt{2}$, represented by the dotted line.

Figure \ref{fig:CEdistrneardielec_Rx_param_epsr_kz_piover128} shows $f_{S_e}(s_e)$ for the incident plus reflected field close to the surface ($k_0z=\pi/128$), at selected values of $\eps / \eps_0$.
 The pdf is seen to make an excursion as $\eps /\eps_0$ increases, returning eventually to the asymptotic $\chi^2_6$ pdf when $\eps \rightarrow +\infty$. 
For larger values of $k_0z$, it has been found that the excursions are much shorter; for example, for $k_0z \sim 2\pi$, there is essentially no longer a discernable parametric dependence of $f_{S_e}(s_e)$ on $\eps_r$.

\section{Effect of conductivity and permittivity on the magnitude of field fluctuations}
Since the probability distribution of each Cartesian component of the underlying complex field $E_\alpha=E^\prime_\alpha - j E^{\prime\prime}_\alpha$ remains circular Gauss normal, on either side of the interface with an isotropic medium, the standard deviation of $E_\alpha$ can be deduced from the $\chi^2_2$ statistics for $|E_\alpha|^2$ or $S_{e_\alpha}$ via
\bea
\sigma_{E_\alpha} = \sqrt{\frac{\langle |E_{\alpha}|^2 \rangle}{2}} = \sqrt{ \frac{\sigma_{|E_{\alpha}|^2}}{2} } = \sqrt{ \frac{ \langle S_{e_\alpha} \rangle }{ \epsilon_{(0)}} }, 
\eea 
where $\alpha=x,y$ or $z$,
with $\langle S_{e_x} \rangle = \langle S_{e_y} \rangle = \langle S_{e_t} \rangle /2$ and $\sigma_{E_x} = \sigma_{E_y} = \sigma_{E_t} / \sqrt{2}$. The dependence of $\sigma_{E_\alpha}$ on $\sigma/(\omega\epsilon_0)$, $\epsilon/\epsilon_0$ and $k_0z$ therefore follows immediately from the results shown in Figs. \ref{fig:CEnearconduc_avgSeCart_param_sigma} and \ref{fig:CEneardielec_Salpha_ifv_epsr}.
Physically, after re-scaling, Fig. \ref{fig:CEneardielec_Salpha_ifv_epsr}(b) indicates that an increased permittivity causes the fluctuations of the refracted fields to exhibit a smaller spread compared to the fluctuations of the incident field, where the latter are quantified by $\sigma_{E_\alpha}$ with $\epsilon/\epsilon_0=1$. This compression of fluctuations is seen to be more prominent for the normal field component than for the tangential field. 
By contrast, Fig. \ref{fig:CEneardielec_Salpha_ifv_epsr}(a) indicates that the spread of the incident plus reflected field increases, on average, with increasing permittivity.

\section{Applications}

\begin{figure}[htb] 
\begin{center} \begin{tabular}{c}
\ \epsfxsize=12cm \epsfbox{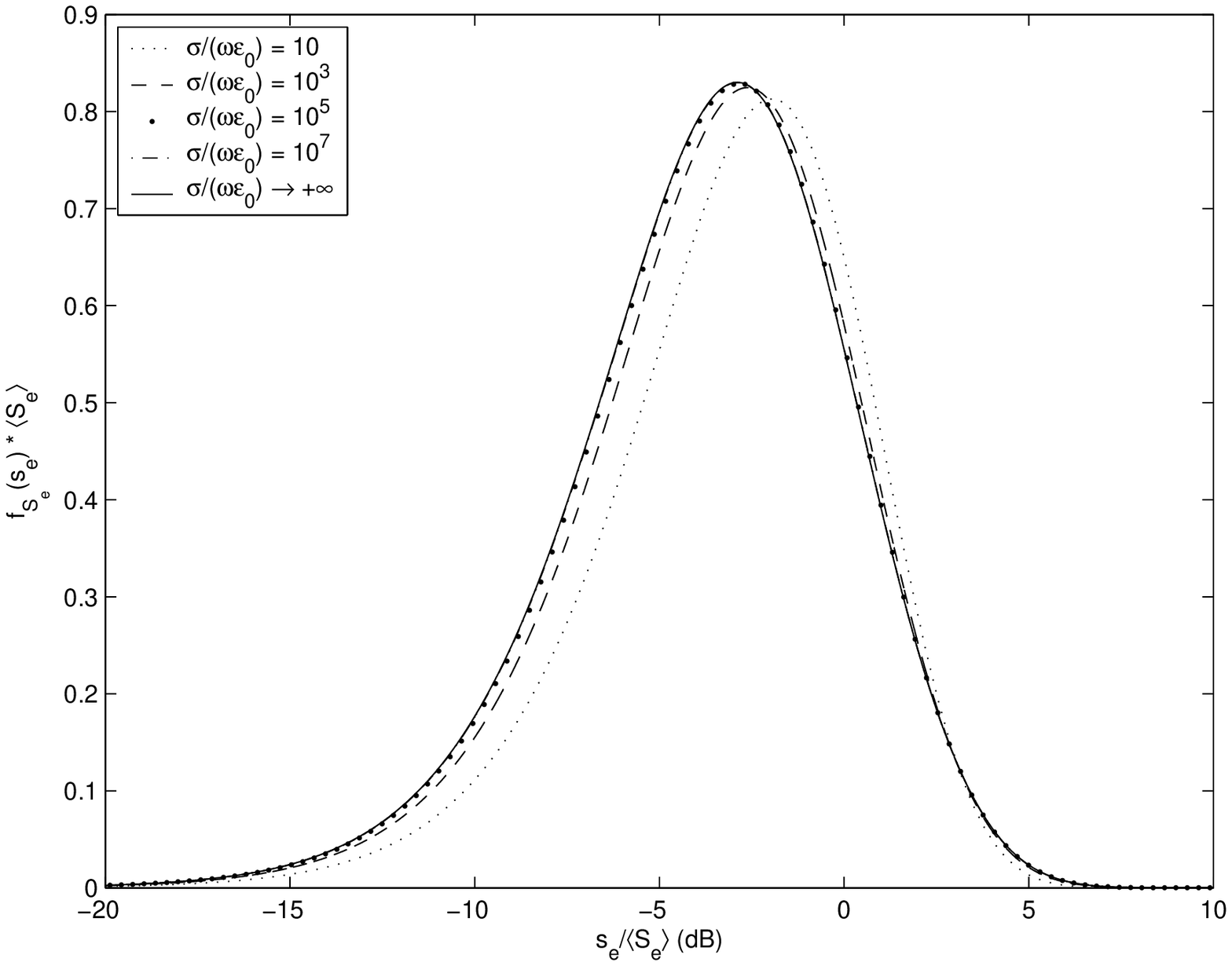}\ \\
\end{tabular}
\end{center}
{\bf \caption{\label{fig:CEdistrnearconduc_param_sigma_kz0_0p7854}
Pdf of $S_e$ of the incident plus reflected field for selected values of $\sigma/(\omega\eps_0)$ at $k_0z=\pi/4$.
}}
\end{figure}

\begin{figure}[htb] 
\begin{center} \begin{tabular}{c}
\ \epsfxsize=12cm \epsfbox{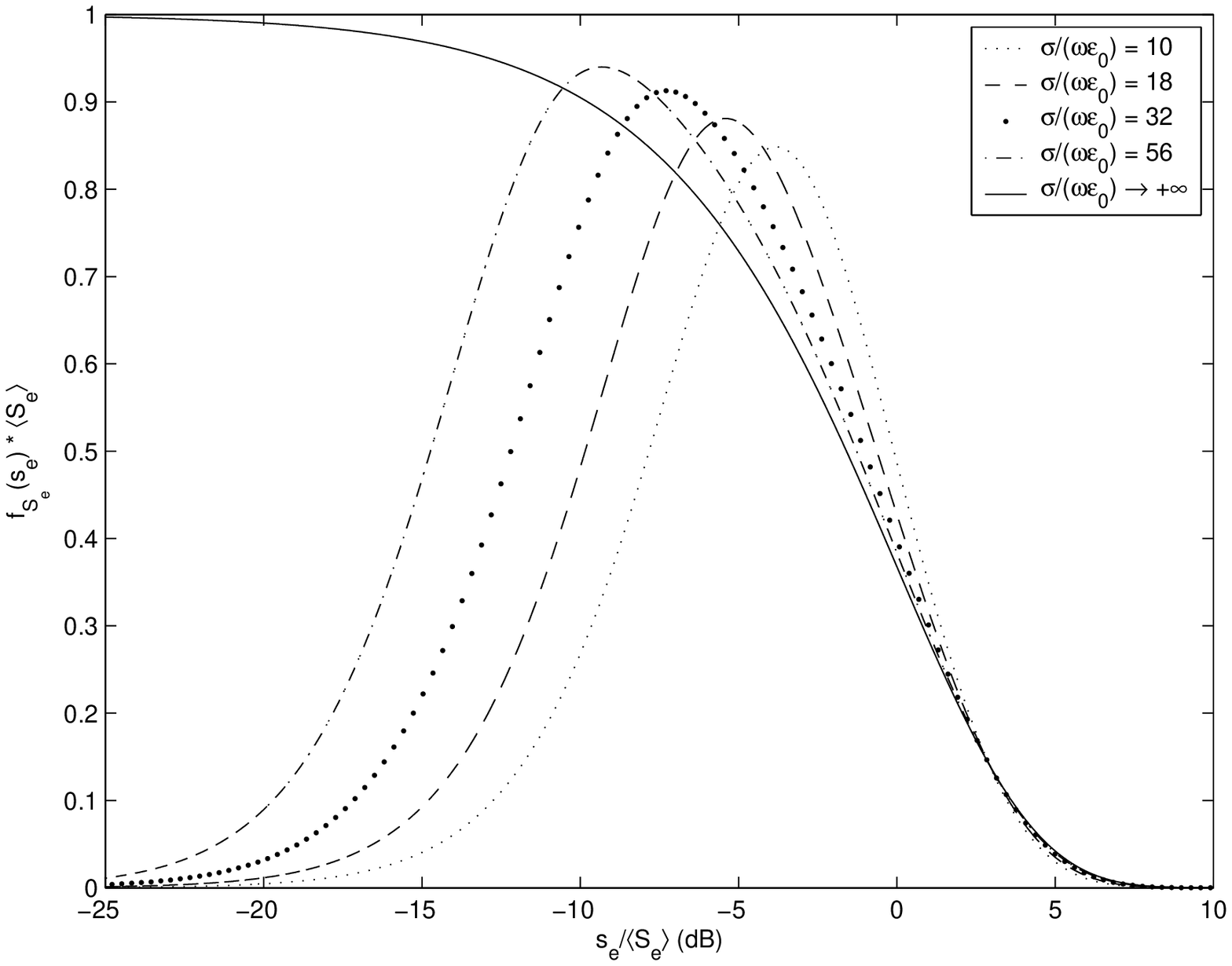}\ \\
\end{tabular}
\end{center}
{\bf \caption{\label{fig:CEdistrnearconduc_param_sigma_kz0_0p01}
Pdf of $S_e$ of the incident plus reflected field for selected values of $\sigma/(\omega\eps_0)$ at $k_0z=0.01$.
}}
\end{figure}

\begin{figure}[htb] 
\begin{center} \begin{tabular}{c}
\ \epsfxsize=12cm \epsfbox{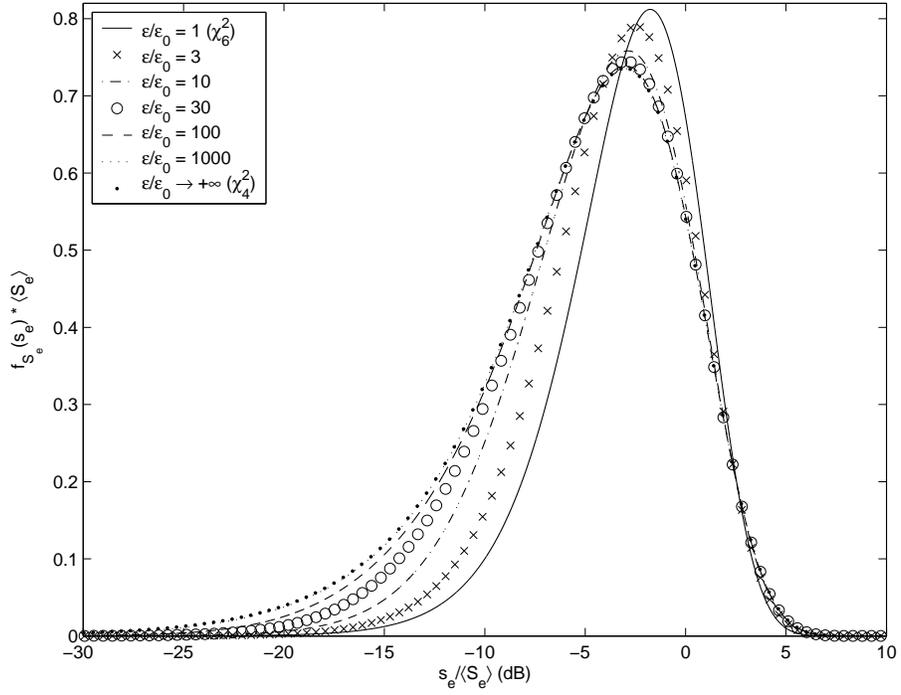}\ \\
\end{tabular}
\end{center}
{\bf \caption{\label{fig:CEdistrneardielec_Tx_param_epsr}
Pdf of electric energy density $S_e$ of the refracted field for selected values of $\eps_r$ at arbitrary $k_0z$.
}}
\end{figure}

\begin{figure}[htb] 
\begin{center} \begin{tabular}{c}
\ \epsfxsize=12cm \epsfbox{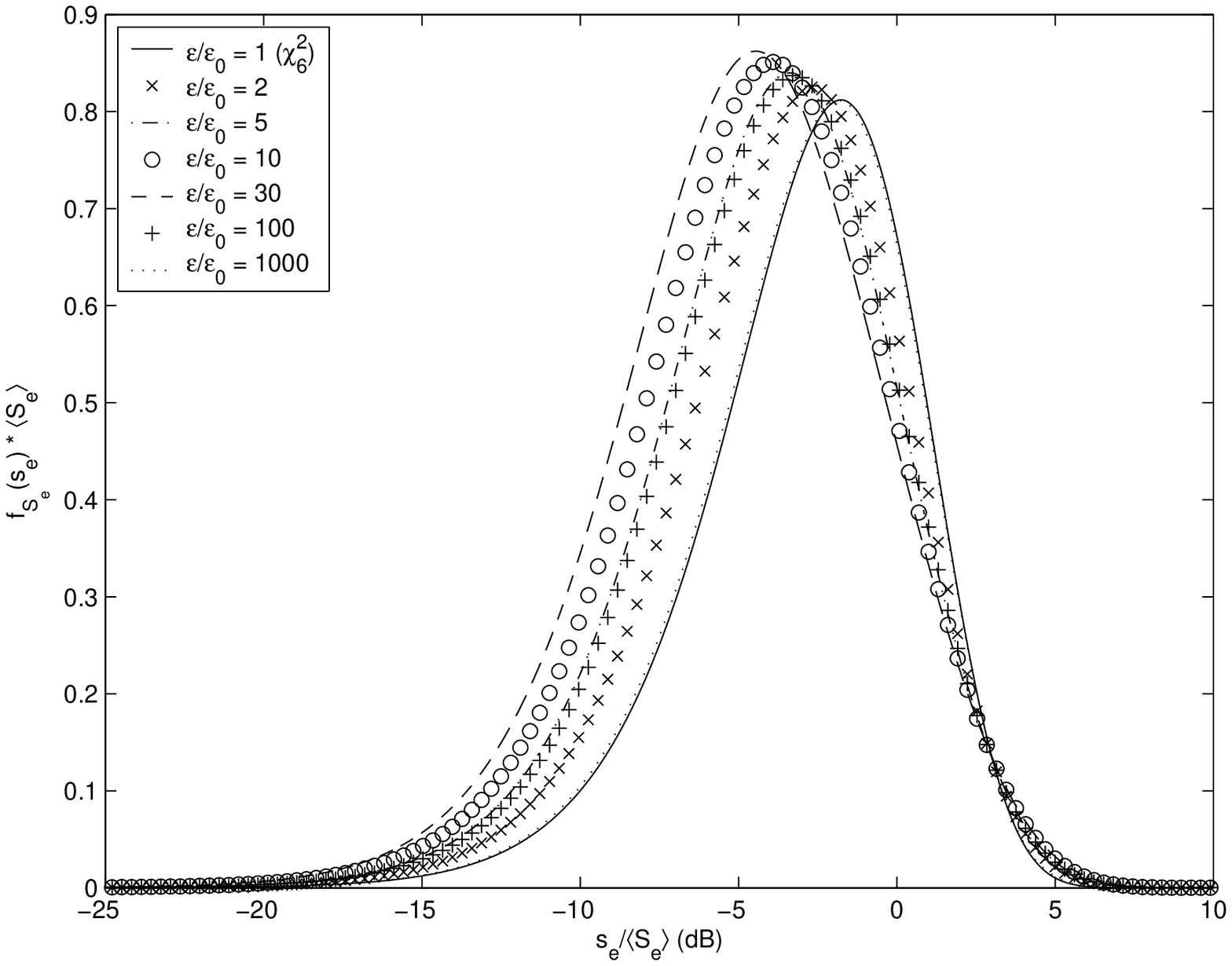}\ \\
\end{tabular}
\end{center}
{\bf \caption{\label{fig:CEdistrneardielec_Rx_param_epsr_kz_piover128}
Pdf of $S_e$ of the incident plus reflected field for selected values of $\eps_r$ at $k_0z=\pi/128$.
}}
\end{figure}

Besides its interest as the solution to a fundamental problem in statistical electromagnetics, the above analysis is relevant to several practical applications, of which we give a examples.
\begin{itemize}
\item {\it Atmospheric and ionospheric propagation of EM waves with applications, including spectroscopy of stellar light:} 
Stellar light propagating through an atmosphere can be represented as a random field radiated by a collection of incoherent point sources that are distributed across one or several narrow solid angle(s) (cf. [Arnaut, 2006b] for a treatment of random fields produced by a spatially filtered EM beam). Upon refraction by atmospheric layers exhibiting permittivities close to $\eps_0$, the light undergoes a change in its distributional and statistical polarization properties; cf. Figs. \ref{fig:CEneardielec_Salpha_ifv_epsr}(b) and \ref{fig:CEneardielec_Pi3_ifv_epsr}(b).
Furthermore, recall that the abovementioned plane-wave expansion refers to a harmonic or quasi-harmonic (narrowband) field for a specific central wavelength. Therefore, the pdfs of light intensities associated with different spectral lines may undergo different changes upon propagation, as a result of frequency dispersion of the medium, thus producing nonuniform changes of the mean values and standard deviations across the received spectrogram. Similar considerations apply to scattered or reflected fields.
In this way, the presented analysis yields corrections that could be instrumental in astrophysical observations and Earth sciences, including remote sensing. 
\item {\it Reflection and refraction of multipath signals for wireless communications:} 
An important issue in the accurate prediction of multipath and wideband propagation through a radio channel is the determination of transfer functions for incident and outgoing signals, and how this transfer might affect the properties (statistical and other) of the received signal. 
By considering the `ratio' of the pdfs in Figs. \ref{fig:CEdistrnearconduc_param_sigma_kz0_0p7854}, \ref{fig:CEdistrnearconduc_param_sigma_kz0_0p01}, \ref{fig:CEdistrneardielec_Tx_param_epsr}, or \ref{fig:CEdistrneardielec_Rx_param_epsr_kz_piover128} 
relative to the $\chi^2_6$ asymptotic distribution of the deep-field energy density, one obtains a transfer function for $f_{S_e}(s_e)$ that captures the change in statistical properties, similar to the way in which Fresnel reflection or refraction coefficients relate incident to reflected or refracted fields. It also yields information on the likelihood and extent of signal distortion.
\item {\it Measurement of constitutive EM properties of materials inside a mode-stirred reverberation chamber, including anisotropy, absorptivity, conductivity and/or permittivity of materials:} 
Several situations in which reverberant fields near a PEC surface are of practical relevance were already mentioned in [Arnaut and West, 2006]; here, we focus on media with finite $\sigma/(\omega\epsilon_0)$ and $\epsilon$ only. Through the measurement of probability distributions of reflected or refracted fields at a fixed and sufficiently close distance from the boundary of a semi-infinite medium or, by extension, from a flat panel of sufficient large electrical thickness, constitutive parameter values can be deduced using Figs. \ref{fig:CEdistrnearconduc_param_sigma_kz0_0p7854}, \ref{fig:CEdistrnearconduc_param_sigma_kz0_0p01}, \ref{fig:CEdistrneardielec_Tx_param_epsr}, \ref{fig:CEdistrneardielec_Rx_param_epsr_kz_piover128} as a statistical inversion problem [Arnaut, 2006b]. Alternatively, the spatial functional $k_0z$-dependence of the average electric energy density (particularly its normal component) can be compared with the results in Figs. \ref{fig:CEnearconduc_avgSeCart_param_sigma} or \ref{fig:CEneardielec_Salpha_ifv_epsr} to deduce these parameters, circumventing the need for determining the pdf. 
\end{itemize}

\section{Conclusion}
In this paper, we have been concerned with the influence of the presence of a deterministic semi-infinite isotropic medium on the local first-order statistical, i.e., distributional properties of an ideal random electromagnetic field for nondirectional (hemispherical) incidence. 
The EM boundary conditions cause statistical anisotropy of the vector field and a redistribution of the energy density between its tangential and normal components, compared to the statistical isotropy and homogeneity of the incident field. This modifies its probability density function. For the reflected plus incident field, this modification changes in a damped oscillatory manner as a function of distance from the interface; for refracted fields, the change is homogeneous within the dielectric because no standing waves occur in this region.
It was found that, for the tangential and normal field components, the presence of a dielectric alters their variability (as expressed by their standard deviation) in an opposing manner, viz., for $\eps/ \eps_0 > 1$ the amplitude fluctuations of the refracted field become more compressed, whereas for the incident plus reflected field the fluctuations expand.

In this paper, the analytical formulation in terms of numerically solved integrals was made possible by the simplicity of the geometry (single interface) and medium (isotropy). More complex configurations involving multiple, finite, angled or curved interfaces, e.g., for interior fields inside overmoded cavities or multiple-scattering problems, are unlikely to afford such an semi-analytical approach. In these cases, one may numerically calculate the statistics by using a Monte Carlo simulation for angular spectra of random plane waves (or other suitable random excitation) with specified input statistics. The numerically solved reflected and refracted fields can then be collated and evaluated at a single location to yield the output statistics for these fields (cf. Sec. V of [Arnaut and West, 2006]).

%
%

\section*{Acknowledgment}
This work was sponsored in part by the 2003--2006 Electrical Programme of the UK Department of Trade and Industry National Measurement System Policy Unit (project no. E03E54).

%
%
%

%
%
%
%
%
%
% ...
%

%
%


\begin{thebibliography}{99}

\bibitem[1]{a}
Arnaut, L. R. (2002), Compound exponential distributions for undermoded reverberation chambers, \textit{IEEE Trans. Electromagn. Compat.}, \textit{44}(3), 442--457.

\bibitem[2]{b}
Arnaut, L. R. and West, P. D. (2006), Electromagnetic reverberation near a perfectly conducting boundary, \textit{IEEE Trans. Electromagn. Compat.}, \textit{48}(2), 359--371.

\bibitem[3]{c} 
Arnaut, L. R. (2006a),
Spatial correlation functions of inhomogeneous random electromagnetic fields,
\textit{Phys. Rev. E}, \textit{73}(3), \# 036604.

\bibitem[4]{d} 
Arnaut, L. R. (2006b),
Spatial correlation functions of random electromagnetic fields in the presence of a semi-infinite isotropic medium,
\textit{Phys. Rev. E}, \textit{74}(5), \# 056610. 

\bibitem[5]{e} 
Booker, H. G. and Clemmow, P. C. (1950), 
The concept of an angular spectrum of plane waves, and its relation to that of polar diagram and aperture distribution, \textit{Proc. Instn. Electr. Engnrs.}, 
\textit{Pt. III}, \textit{97}, 11--17.

\bibitem[6]{f}
Bourret, R. C. (1960), Coherence properties of blackbody radiation, \textit {Nuovo Cimento}, \textit {XVIII}(2), 347--356.

\bibitem[7]{g}
Dunn, J. M. (1990), Local, high-frequency analysis of the fields in a mode-stirred chamber, \textit{IEEE Trans. Electromagn. Compat.}, \textit{32}(1), 53--58.

\bibitem[8]{h} 
Eckhardt, B., D\"{o}rr, U., Kuhl, U., and St\"{o}ckmann, H.-J.   (1999),
Correlations of electromagnetic fields in chaotic cavities, 
\textit{Europhys. Lett.}, \textit{46}(2), 134--140.

\bibitem[9]{k} 
Hill, D. A. (1998)
Plane wave integral representation for fields in reverberation chambers, 
\textit{IEEE Trans. Electromagn. Compat.}, \textit{40}(3), 209--217.

\bibitem[10]{l} 
Hill, D. A. and Ladbury, J. M. (2002),
Spatial-correlation functions of fields and energy density in a reverberation chamber, 
\textit {IEEE Trans. Electromagn. Compat.}, \textit{44}(2), 95--101.

\bibitem[11]{m} 
Mehta, C. L. and Wolf, E. (1964),
Coherence properties of blackbody radiation. I. Correlation tensors of the classical field, 
\textit {Phys. Rev.}, \textit{134}(5A), A1143--A1149.

\bibitem[12]{n}
Sarfatt, J. (1963),
Quantum-mechanical correlation theory of electromagnetic fields, 
\textit {Nuovo Cimento}, \textit{XXVII}(5), 1119--1129.

\bibitem[13]{o} 
Whittaker, E. T. (1902),
On the partial differential equations in mathematical physics, \textit{Math. Ann.}, \textit{LVII}, 333--355, \S\S 5.1, 5.2.
\end{thebibliography}
\end{document}